\documentclass[useAMS,usenatbib]{mn2e}
\usepackage{psfig}

\usepackage{times}

\def\lesssim{{_ <\atop{^\sim}}}
\def\grtsim{{_ >\atop{^\sim}}}

\def\fb{\mbox{$f_{\rm b,gal}$}}
\def\fbuniv{\mbox{$f_{\rm b,Univ}$}}

\def\kms{\mbox{kms$^{-1}$}}
\def\vesc{\mbox{$V_{\rm esc}$}}

\def\ms{\mbox{$M_{\rm *}$}}
\def\mh{\mbox{$M_{\rm h}$}}
\def\msun{\mbox{M$_\odot$}}
\def\Mso{\mbox{$M_{\rm pcg}$}}
\def\Msop{\mbox{$M_{\rm pcg}^{\prime}$}}

\def\vm{\mbox{$V_{\rm m}$}}

\def\LCDM{\mbox{$\Lambda$CDM}}

\def\apj{\mbox{ApJ}}
\def\apjs{\mbox{ApJSS}}
\def\mnras{\mbox{MNRAS}}
\def\apjl{\mbox{ApJL}}

\def\aap{\mbox{A\&A}}
\def\aj{\mbox{AJ}}

\def\spose#1{\hbox to 0pt{#1\hss}}
\newcommand\lsim{\mathrel{\spose{\lower 3pt\hbox{$\mathchar"218$}}
     \raise 2.0pt\hbox{$\mathchar"13C$}}}
\newcommand\gsim{\mathrel{\spose{\lower 3pt\hbox{$\mathchar"218$}}
     \raise 2.0pt\hbox{$\mathchar"13E$}}}

\title[Galaxy outflows and re--accretion]
{Can galaxy outflows and re--accretion produce the downsizing in specific star formation 
rate of late--type galaxies?}

\author[Firmani et al.]
{C. Firmani$^{1,2}$\thanks{E--mail: firmani@merate.mi.astro.it}, 
V. Avila-Reese$^{2}$\thanks{E--mail: avila@astro.unam.mx}, A. Rodr\'{\i}guez-Puebla$^{2}$\\
$^{1}$Osservatorio Astronomico di Brera, via E.Bianchi 46, I-23807
Merate, Italy\\
$^{2}$Instituto de Astronom\'{\i}a, Universidad Nacional Aut\'onoma de M\'exico,
A.P. 70-264, 04510, M\'exico, D.F.}

\begin{document}


\pagerange{\pageref{firstpage}--\pageref{lastpage}} \pubyear{2002}

\maketitle

\label{firstpage}

\begin{abstract}

An increasing amount of recent observational evidence shows that
less massive galaxies are, the higher on average is their specific star formation rates 
(SSFR = SFR/\ms, \ms\ is the stellar mass). Such a trend, called the 'SSFR downsizing'
(SSFR--DS) phenomenon, is seen for local and high--redshift (back to $z\sim 1-2$) 
galaxy samples. We use observational data related only to {\it disc} galaxies and 
explore how does the average SSFR change with $z$ for different masses. For all the masses
in the range $\sim 10^{9.5}-10^{10.5}$ \msun, the SSFR increases with ($1+z$) to a 
power that seems not to depend on \ms, and at all redshifts smaller galaxies have ever
higher SSFRs; galaxies less massive than $\ms\sim 10^{10}$ \msun\ are forming stars
at a greater rate than in the past assuming constant SFRs over a Hubble time to build
stellar mass. 
We show that these features strongly
disagree with the $\Lambda$ Cold Dark Matter (\LCDM) halo hierarchical mass accretion
rates. Further, by means of self--consistent models of disc galaxy evolution inside growing 
\LCDM\ halos, the effects that disc feedback--driven outflows and gas 
re--accretion have on the galaxy SSFR histories are explored.
The parameters of the outflow and re--accretion schemes are tuned to reproduce the
present--day \mh--\ms\ relation (\mh\ is the halo mass) inferred from the observationally--based
\ms\ function of disc galaxies. In the case of outflows only, the SSFR of individual model 
galaxies increases with $z$ roughly as 
($1+z$)$^{2.2}$ for all the masses (somewhat shallower than observations) with a 
normalization factor that depends on mass as $M_*^{0.1}$, i.e more 
massive galaxies have slightly larger SSFRs, contrary to the observed strong SSFR--DS 
trend. For the re--accretion cases, the dependence on $z$ remains approximately the same
as without gas re--infall, but the correlation on mass increases even for most of the 
reasonable values of the model parameters. The comparison of models and observations
in the SSFR--\ms\ plane at $z\sim 0$ (where the data are more reliable), clearly
shows the divergent trend in SSFR as the masses are lower (upsizing vs downsizing). 
We explain why our models show the reported trends, and conclude that the SSFR--DS 
phenomenon for low--mass galaxies poses a sharp challenge for \LCDM--based disc 
galaxy evolution models.  

\end{abstract}

\begin{keywords}
cosmology: theory --- galaxies: evolution --- galaxies: haloes --- galaxies: high-redshift 
--- galaxies: spiral --- galaxies: star formation
\end{keywords}

\section{Introduction}

The inference of the assembling history of stellar populations in galaxies as
a function of their morphological type and luminosity (mass) is 
a major topic of present--day extragalactic astronomy, as well as a 
key probe for cosmologically--based models of galaxy formation and evolution.

Based on chemical and spectro-photometric studies 
of local galaxies, it has long been known that more massive/earlier type galaxies 
hold on average older stellar populations and were formed over a shorter time-span, than 
less massive/later type galaxies (e.g., Faber et al. 1992; Worthey et al. 1992;
Carollo, Danziger \& Buson 1993; Bell \& de Jong 2000). With the advent of large and complete
surveys of local galaxies and the significant improvement of the population synthesis 
models, the average ages, and even the whole star formation rate (SFR) histories of all kinds
of galaxies, have been inferred (e.g., Kauffman et al. 2003; Heavens et al. 2004; 
Jimenez et al. 2005; Gallazzi et al. 2005,2008; Cid Fernandez et al. 2005; Panter et al. 2007). 
The general conclusion of these works confirms and extends the previous results: 
more massive galaxies assembled their stars at earlier epochs and on shorter time 
scales, halting their star formation (SF) at later epochs. This phenomenon has been now
dubbed as {\it 'archaeological downsizing'} (Thomas et al. 2005; Fontanot et al. 2009).

In the last decade, the observational capabilities allowed the extension
of galaxy population studies to high redshifts. With these 'look-back
studies', the current properties of galaxies at their observed times, as well as their
past histories, were inferred.
In a pioneering work, Cowie et al. (1996) have found that the maximum rest--frame $K-$band
luminosity of actively star--forming galaxies has declined with time in the 
redshift range $z=1.7-0.2$, i.e. the SF efficiency shifts to lower mass galaxies 
with time. This phenomenon has been
confirmed by subsequent works based on wider surveys with multi-wavelenght information
(Juneau et al. 2005; Feulner et al 2005; Bell et al. 2007; Noeske et al. 2007a,b; 
Zheng et al. 2007; Cowie \& Berger 2008; Mobasher et al. 2009; Chen et al. 2009; Damen et al.  2009; 
see Fontanot et al. 2009 for more references). These works allowed to infer the specific SF rates 
(SSFR = SFR/\ms, \ms\ is the galaxy stellar mass) for relatively complete samples of 
galaxies at different redshifts until $z\sim 1-2$. The general result is that the measured
SSFRs tend to be higher as smaller is \ms, a phenomenon called {\it `SSFR downsizing'}, this trend 
being observed at all the studied redshifts.
Recent look--back studies of galaxy luminosity functions have also confirmed the local 
inferences of archaeological downsizing mentioned above: the high--mass end of the 
galaxy mass function seems to be mostly in place since $z\sim 2$, while the abundances 
of galaxies of smaller stellar masses grow gradually with time (Daddi et al. 2004,2007; 
Bundy et al. 2004,2006; Drory et al. 2005; Conselice et al. 2007; Marchesini et al. 2009;
P\'erez-Gonz\'alez et al. 2008 and see more references therein).

The emerging observational picture of downsizing in the stellar mass galaxy assembly
has been confronted with the hierarchical clustering
scenario of galaxy formation and evolution, based on the $\Lambda$ Cold
Dark Matter (\LCDM) cosmological model. At this point, as shown in
Neistein et al. (2006; see also Fontanot et al. 2009), it is important to realize that 
the downsizing effect to which observations refer generically, has actually many 
manifestations, related to different phenomena, involving different types of galaxies
and different epochs in their histories. Herein we emphasize at least the two
distinct downsizing phenomena mentioned above: 
\begin{enumerate}
\item the {\it 'archaeological downsizing'}
related to the early ($z\grtsim 2$) assembly of most stars in massive/early 
type galaxies, and 
\item the {\it 'SSFR downsizing'} (hereafter SSFR--DS) related to the later growth 
of the relative stellar mass as less massive are the galaxies.
\end{enumerate}

The archaeological downsizing has a partial explanation in the frame of the hierarchical 
clustering process of dark--matter halos (Mouri \& Taniguchi 2005; Neistein et al. 2006; 
Guo \& White 2008; Kere{\v s} et al. 2009), and {\it astrophysical} processes like Active 
Galactic Nuclei (AGN)--feedback have been introduced also to get better agreement with 
the observations (e.g., Bower et al. 2006; Croton et al. 2006; Cattaneo et al. 2006; 
Monaco, Fontanot \& Taffoni 2007; Hopkins et al. 2008; Somerville et al. 2008, 
and see more references therein). 

Regarding the SSFR--DS, which is the focus of this work, if confirmed, it will turn as 
a pervasive problem for current models. This problem refers mainly to the high SSFRs of 
low--mass ($\ms\sim 10^9-3\times 10^{10}\msun$) star--forming disc galaxies. For these 
galaxies the AGN feedback is not expected to be important because either they do not
have AGNs (e.g., Kauffmann et al. 2003; Salim et al. 2007) or their AGNs are too weak
as to produce significant feedback (e.g., Bower et al. 2006; Crotton et al. 2006). By using 
semi--analytical models (SAMs) several authors 
have found systematically that the SF in modeled low--mass galaxies happens too early 
and is over--quenched at later times, showing these galaxies have too low SSFRs at low redshifts
as compared with observational inferences (see for recent results and more references 
Somerville et al. 2008 and Fontanot et al. 2009). 
The problem is sharpened by the fact that in the SAMs, the baryon mass fraction
of low--mass galaxies has to be decreased systematically
in order to reproduce the low--luminosity 
end of the luminosity function (e.g., Benson et al. 2003). This is accomplished by 
introducing appropriately tuned schemes of very strong stellar ejective feedback as a 
function of mass, which worsens the SSFR--DS problem (\S 4).  
Late re--infall of the ejected gas (Bertone, De Lucia \& Thomas 2007; Oppenheimer \& Dav\'e 2008)
could offer a partial cure to this problem.

In this work we aim to 'isolate' the SSFR--DS problem for only low--mass disc 
galaxies and explore in a transparent way the effects of ejective stellar feedback and 
gas re--accretion on the SSFR histories of galaxies by means of evolutionary
hydrodynamical models of disc galaxies formed inside growing CDM halos.  
We will show that a solution to the SSFR--DS problem is not simple in the context
of the hierarchical \LCDM\ scenario.     

In Section 2, the observational data to be used in this paper are presented and discussed.
Preliminary theoretical predictions regarding the SSFR histories of galaxies as a function
of mass are presented in Section 3.
A brief review of our galaxy disc evolutionary models and the parametric schemes introduced
for modeling ejective feedback and gas re--infall are given in Section 4. The model results
and their comparison to observations are presented in Section 5. Finally, a summary
of the results and a discussion are presented in Section 6.  

The cosmological model used  throughout this paper is the concordance one with $h=0.7$, 
$\Omega_{\Lambda}=0.7$, $\Omega_m=0.3$, $\Omega_b=0.042$, and $\sigma_8=0.8$.

\section{The observations}

\begin{figure}
\vspace{8.8cm}
\includegraphics{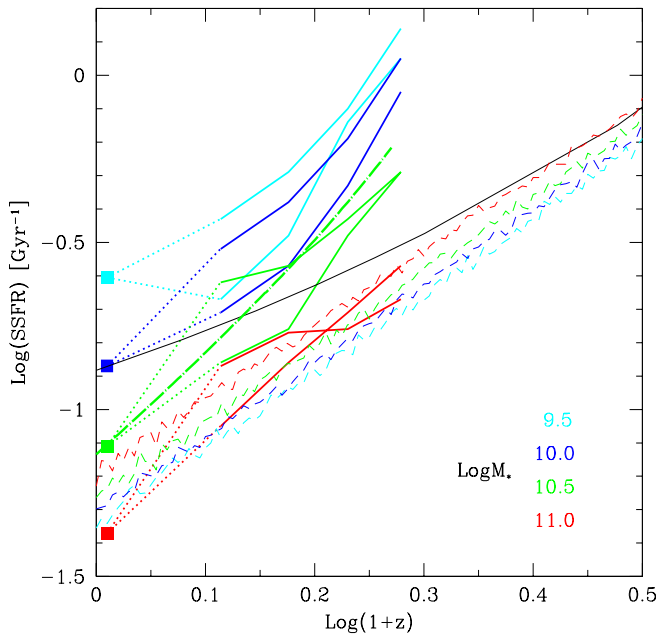}
\caption{Evolution of the SSFR of late--type galaxies for different fixed values of \ms. 
The data at $z\sim 0$ (squared dots) are from Salim et al. (2007). 
The solid curves were inferred from Bell et al. (2007) for their two fields,
CDFS and A901 (corresponding upper and lower curves, respectively; see text). 
From top to bottom, the data correspond to mass bins centered at Log(\ms/\msun) =
9.5 (cyan), 10.0 (blue), 10.5 (green), and 11.0 (red), respectively. The thin
dot--long--dashed green line is our estimate from the data for
the individual evolution of an average galaxy that ends at $z=0$ with
$\ms = 10^{10.5} \msun$ (see text).  The solid black line shows the
curve $1/[t_H$($z$) -- 1 Gyr](1-R) corresponding to constant SFR along
the time.  The dashed curves are 'dark--matter halo' SSFR histories calculated
as $\dot{\mh}/[\mh\ (1-R)]$ (see text); from top to bottom the 
corresponding masses are $\mh=10^{12.6} \msun$ (red), $\mh=10^{12.1} \msun$ (green), 
$\mh=10^{11.7} \msun$ (blue), and $\mh=10^{11.4} \msun$ (cyan). The corresponding
stellar masses are roughly as those of the data. The cosmological specific infall
rates drive SSFRs that are slightly higher for massive galaxies and strongly lower
for low--massive galaxies than observational inferences.
}
\label{obs}
\end{figure}


Our aim is to target the potential problem of SSFR--DS, and then compare 
observations with detailed theoretical predictions.  We focus our study here only 
on normal (not dwarf) low--mass {\it disc} galaxies ($10^{9.5}\ms\lesssim 10^{10.5}$ \msun).
On one hand, due to completeness issues, dwarf galaxies are not taken into account in the 
$z\sim 0$ and high--redshift observational samples to be used here. On the other hand, 
since most of dwarf galaxies are satellites in bigger systems, the physics of them is affected
 by several environmental processes (ram pressure, tidal stripping, interactions,
etc.), which are not considered in our models.

In order to determine the SSFR--\ms\ relation at different redshifts, multi--wavelength
photometric and spectroscopic extensive samples are necessary. The stellar masses and SFRs 
of galaxies are inferred commonly by using fitting techniques to stellar population synthesis 
(SPS) models; the SFRs of galaxies can be inferred also by using emission lines that trace 
SF when available. The uncertainties associated to these inferences are large due
to selection effects, sample incompleteness, and the methods used to infer \ms\ and  the
SFR (see for a discussion \S\S 6.1). 
For our work, we require samples where late--type galaxies are separated from early--type ones.
This is the case of the observational works by Salim et al. (2007; $z\sim 0$) and 
by Bell et al. (2007; high redshifts). 

Bell et al. (2007; see also Zheng et al. 2007 and Damen et al. 2009) analyze the COMBO-17
photometric redshift survey combined with {\it Spitzer} $24\mu$m data ($0.2<z\lesssim 1.0$). 
Stellar masses were estimated using 17 pass-bands in conjunction with a non-evolving template 
library derived from the P\'egase SPS model, and SFRs were determined from the combined UV 
and IR fluxes of galaxies. In Fig. \ref{obs}, we reproduce the averaged data presented 
in their Fig. 3 (SSFR vs. \ms), where four 
redshift bins centered on $z\sim 0.9$, 0.7, 0.5, and 0.3 are given. 
We reproduce the data associated only with the blue cloud (late--type) galaxies and for four 
masses [Log(\ms/\msun) = 9.5, 10.0, 10.5, and 11.0 from top to bottom, respectively]. 
Their data are separated according to two COMBO-17 fields observed by {\it Spitzer}: 
the extended Chandra Deep Field South (CDFS), and the field around the Abell 901/902 (A901) 
galaxy cluster (upper and lower curves in Fig. \ref{obs}, respectively). 
Note that the SSFRs from CDFS are systematically higher than those around the denser field of 
a cluster. 
It is well known that the SF activity tends to be quenched in galaxies in high--density regions. 
Thus, for comparison with our theoretical predictions, the data corresponding to CDFS 
are likely more appropriate.

The data in Fig. \ref{obs} corresponding to $z\sim 0$ were taken from Salim et al. (2007), who
obtained \ms\ and dust--corrected SFRs for $\approx 50,000$ galaxies by fitting the SDSS
and $GALEX$ photometry to a library of dust--attenuated SPS models. They were able to 
separate from their sample the ``pure'' star--forming galaxies with no AGN, which form 
a well defined linear sequence in the SSFR vs. \ms\ plot, fitted linearly by 
Log SSFR = --0.35(log \ms -- 10) -- 9.83. In Fig. \ref{obs} the same four masses 
related to the data from Bell et al. (2007) are used for the 
Salim et al. (2007) results at $z\sim 0$.

From Fig. \ref{obs} we see that the SSFR of what roughly can be considered late--type galaxies 
declines from $z\sim 1$ to $z\sim 0$ almost with the same slope for all the masses. However, 
the normalization strongly depends on \ms, {\it being on average the SSFRs at all
redshifts higher as lower is \ms.}  
This shows the SSFR-DS behavior. A crude approximation to the range of the data 
displayed in Fig. \ref{obs} is:
\begin{equation}
SSFR = 0.13 \ M_{*,10}^{-0.53}
\ (1+z)^{2.6} \ {\rm Gyr^{-1}},
\label{SSFRint}
\end{equation}
where $M_{*,10}$ is \ms\ in units of $10^{10} \msun$. 
Recently, for all (late-- and early--type) galaxies and back to $z\approx 2.8$,
Damen et al. (2009) have reported a trend of SSFR on redshift also not strongly depending on \ms,
with a trend roughly proportional to $(1+z)^5$ (see also Feulner et al. 2005; Zheng et al. 
2007; Martin et al. 2007; P\'erez-Gonz\'alez et al. 2008). There are some pieces of 
evidence that the SSFRs of massive 
galaxies start to increase significantly and overcomes the SSFRs of less massive galaxies 
only at redshifts larger than 2 (P\'erez-Gonz\'alez et al. 2008)
   
Notice that  each of the pairs of curves inferred from observations for $z\grtsim 0.3$
in Fig. \ref{obs} refer to a given \ms\ that is the same at each redshift, i.e. these curves 
do not refer to the evolution of individual galaxies (evolutionary tracks), but to galaxy 
populations at different epochs. 
We have carried out an exercise for recovering approximate galaxy evolutionary tracks 
from the observational data in diagram of Fig. \ref{obs}. 
We assume that a galaxy of mass \ms\ at redshift $z$ increases its mass during the 
time interval $(dt/dz)dz$ by forming stars at a rate $\dot{\ms} = (1-R) SSFR(\ms,z) \ms(z)$, 
where $SSFR$ is given by eq. (\ref{SSFRint}), and $R=0.4$ takes into account for gas return 
due to stellar mass loss. This way, we are able to calculate $\dot{\ms} / \ms$ at each
$z$ for an ``average'' galaxy evolutionary track that ends at $z\sim 0$ with a given \ms.
The dot-dashed line shows the average evolutionary track for a present--day galaxy stellar 
mass of $10^{10.5} M_\odot$. The evolutionary tracks in the range of the data are roughly 
parallel and at $z \sim 0$ intercept the SSFR axis at the level of the corresponding masses.
As seen in Fig. \ref{obs}, the evolutionary track inferred this way is only slightly steeper 
than the observed constant mass curves.

\section{Preliminary theoretical predictions}

From a theoretical point of view, in order to estimate the level of SF activity of
galaxies as a function of their masses and epochs, let consider first the case of a 
constant SFR history. Then, \ms\ at the cosmic time $t_H$($z$) should be given by
$\ms\sim $ SFR $\times$(1 -- R)$\times $($t_H$($z$) -- 1 Gyr); 1 Gyr is subtracted 
in order to take into account the (average) delay in the formation
of halos that host galaxies. 
If the measured SSFR at $z=0$ is smaller (larger) than $\sim 1/[t_H $($z$) -- 1 Gyr] (1--R), 
then the average SFR of the given galaxy has been higher (lower) in the past than in the 
current epoch. In Fig. \ref{obs} the curve $1/[t_H$($z$) -- 1 Gyr](1-R) for the cosmology 
used here and for R=0.4 is plotted (thin solid line).
The constant SFR case is just indicative of the situation: low--mass galaxies ($\ms\lesssim
10^{10} \msun$) show current SFRs higher than their past average SFR. 

Now, within the context of the \LCDM\ cosmogony, one may anticipate a potential disagreement with observations related to the SSFR--DS phenomenon. 
By using an extended Press--Schechter approach (Avila-Reese et 
al. 1998; Firmani \& Avila-Reese 2000 --hereafter FA2000; see \S 4), we calculate for a given 
present--day \mh\ its averaged mass aggregation 
history (MAH), $\mh(z)$, from tens of thousands of Monte-Carlo extractions. By assuming 
that the SSFR is driven by the halo specific mass aggregation rate, $(\dot{\mh}/\mh)(z)$, 
and correcting by the gas return factor (1 -- R), we have calculated the 'dark matter'--driven 
SSFR histories for different halo masses. In Fig. \ref{obs} the dashed lines show from top 
to bottom the SSFR histories for 
halo masses from $\mh=10^{12.6} \msun$ (red) to $\mh=10^{11.4} \msun$ (cyan), which 
correspond roughly to stellar masses from $\ms=10^{11} \msun$ to $\ms=10^{9.5} \msun$, 
respectively (see \S\S 4.1 and Fig. 2).  A good interpolating formula to our results 
since $z\sim 1$ is:
\begin{equation}
SSFR\approx \frac{1}{(1-R)} \frac{\dot{\mh}}{\mh}
= \frac{ 10^{-1.5} }{(1-R)} M_{*,12}^{1/8} 
\ (1+z)^{2.25} \ {\rm Gyr^{-1}}
\label{SSFRdm}
\end{equation}
where $M_{*,12}$ is \ms\ in units of $10^{12} \msun$.

From this very preliminary calculation, we conclude that while the dark matter--driven 
SSFR histories of massive galaxies do not show a significant discrepancy with observations, for 
low--mass galaxies ($\ms\lesssim 10^{10.5}\msun$) the discrepancy becomes significant 
and it increases as smaller is the mass. The predicted SSFRs of low--mass galaxies are
lower than the observed ones beyond the uncertainty level. A first guess to solve
such a conflict within the context of the hierarchical cosmogony is to propose 
that the main stellar formation epoch of low mass galaxies is delayed by some 
astrophysical processes to lower redshifts. Can gas outflows and later re--accretion work
in this direction? In what follows, we turn out to self--consistent models of disk galaxy
formation and evolution to explore this question.

\section{The model}

The formation and evolution of disc galaxies within growing dark matter halos can be followed 
in a transparent and {\it self-consistent} way by using simplified hydrodynamic models of discs 
in centrifugal and vertical hydrostatic equilibrium, and with a SF mechanism triggered by 
a disc instability criterion and self--regulated by an energy 
balance process (the {\it semi-numerical} approach; FA2000; Avila-Reese \& Firmani 2000; see for
similar approaches van den Bosch 2000; Naab \& Ostriker 2006; 
Stringer \& Benson 2007; Dutton \& van den Bosch 2009 --hereafter DvdB09). 
 
The main physical ingredients of the models used here are as follows. 
A special extended Press--Schechter approach based on the conditional probability
(Lacey \& Cole 1993) is used to generate the halo mass accretion 
histories (MAHs) from the primordial Gaussian density fluctuation field. 
A generalized secondary infall model with elliptical orbits is applied to calculate 
the time--by--time virialization of the accreting mass shells (Avila-Reese et al. 1998).
The orbit ellipticity parameter is fixed in such a way that the structure of the \LCDM\ 
haloes agrees with results from cosmological N--body simulations 
(Avila-Reese et al. 1999; FA2000). A (baryon) fraction of the mass of each accreting 
shell is assumed to cool down in a dynamical time and 
form a disc layer\footnote{For halo masses lower than $\sim 5\ 10^{11}$ \msun\ 
the radiative cooling time is ever shorter than the Hubble time, $t_H$. 
Therefore, this assumption works well for halos of this mass and smaller. 
For larger masses, the cooling time becomes larger than $t_H$. 
Then, the mass baryon fraction available to form the galaxy decreases systematically  with mass.}. 
We call 'primary' cosmological gas accretion that one related to these
accreting mass shells with an universal baryon fraction $\fbuniv\equiv \Omega_B/\Omega_{DM}
=0.163$ for the cosmology adopted here. The final galaxy baryon fraction, 
$\fb\equiv M_{\rm gal}/\mh$, is determined by the further disc gas ejecting and 
infalling processes.

The accreting mass shells at the time of their virialization, $t_v$, are assumed to rotate 
rigidly  with a specific angular momentum calculated as $j_{sh}(t_v)=\Delta J_h/\Delta M_h$, 
where $\Delta$ represents a difference between two time steps, and 
$J_h=\lambda _h GM_h^{5/2}/\left| E_h\right| ^{1/2}$, $M_h$, and $E_h$, are the halo 
total angular momentum, mass, and energy respectively; $\lambda _h$ is the halo spin 
parameter, assumed to be {\it constant in time}.
As a result of the assembling of these mass shells, a present--day halo ends with an angular 
momentum distribution close to the (universal) distribution measured in N-body simulations 
(Bullock et al. 2001). The radial mass distribution of the layer is calculated by equating 
its specific angular momentum to that of its final circular orbit in centrifugal equilibrium 
(detailed angular momentum conservation). The superposition of these layers form a 
gaseous disc, which tends to be steeper in the centre and flatter at the periphery than 
the exponential law. The gravitational interaction of disc and inner halo during their 
assembly is calculated using the adiabatic invariance formalism.

A further step is the calculation of the {\it stellar} surface density profile. 
The previously mentioned processes and the fact that SF is less efficient at 
the periphery than in the centre produces stellar discs with a nearly exponential 
surface density distribution, and the size mainly determined by $\lambda _h$.  
The disc SF at a given radius (assuming azimuthal symmetry) is triggered by the Toomre 
gas gravitational instability criterion and self--regulated by a balance between the 
energy input due to SNe and the turbulent energy dissipation in the ISM. 
Both ingredients determine the gas disc height and the SF rate. This physical 
prescription naturally yields a Schmidt-Kennicutt-like law. The SF efficiency depends 
on the gas surface density determined mainly by $\lambda _h$, and on the gas infall 
history, which in its primary phase is proportional to the halo MAH.


\begin{figure}
\vspace{13.8 cm}
\includegraphics{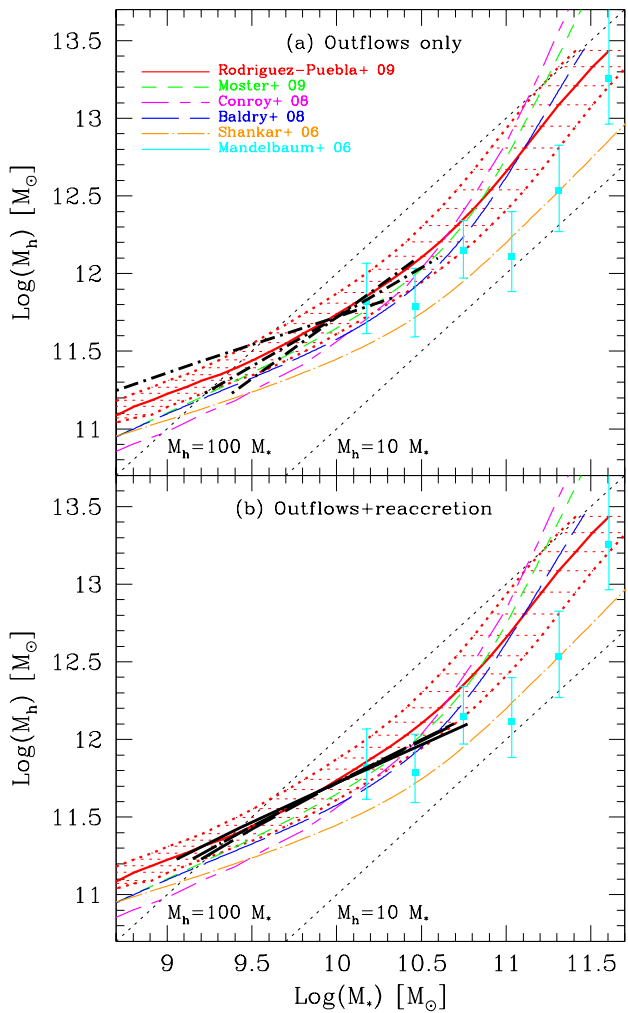}
\caption
{{\it Panel (a):} Semi--empirical and modeled stellar vs halo masses. The former are inferred 
by different authors by matching a given observed luminosity (\ms) function with the 
theoretical halo mass function for all galaxies and halos (non-continuous curves), for 
only late--type galaxies and halos that likely host late--type galaxies (continuous red 
curve for the average, and dotted curves encompassing the dotted horizontal lines for the 
1 $\sigma$ uncertainty), and from direct weak gravitational lensing studies (cyan solid 
squares with vertical error bars). The corresponding literature sources are indicated 
inside the figure. For the model predictions (thick dot--dashed curves), the disc outflow 
cases for $n = 1,$ 2, and 5 (respectively in clockwise sense) are shown. The models
are forced to agree with the continuous red line at $10^{10}$ \msun\ by tuning the free
 parameter $\epsilon$ in eq. (\ref{outflow}); see for the values Table 1. The $n=2$ 
'energy--driven' outflow case best reproduces the observational inferences.
{\it Panel (b):} Same as panel (a) but for models that include gas re--accretion. Only
cases with $n=1$ and \Mso=const. ($\mu = 0$, see eq. \ref{reinfall}) are plotted.
For larger values of $n$, the re--accretion produces too shallow slopes in the \mh--\ms\
relation. The three cases plotted with thick solid, dashed, and dot--dashed lines, 
correspond to values for the (Log($m$), $\alpha$) parameters of (10.0, 1.0),
(10.5, 1.0), and (10.0, 0.5), respectively (see Table 1).
}
\label{MhMs}
\end{figure}


\subsection{Feedback--driven outflows}

In our previous works, we did not take into account disc mass outflows 
(galactic super-winds) due to SN and radiation pressure feedback. 
Therefore, the driving factor for disc growth was solely the gas infall rate assumed 
proportional to the halo mass aggregation rate, where the proportionality coefficient 
is the galaxy baryon mass fraction, \fb. 
This fraction was fixed to a constant value smaller than the universal baryon fraction, 
\fbuniv, by a factor $\sim 4$. 
With this simple assumption, nearly flat rotation curves at the present and at higher 
redshifts, and good agreement with observed disc scaling relations were obtained 
(e.g., FA00; Avila-Reese et al. 2008; Firmani \& Avila-Reese 2009).

Here we introduce an algorithm for outflows from the disc similar to those used 
in the SAMs but applied {\it locally} to the evolving disc--halo system as a function of the 
radius (DvdB09). The outflow is assumed to move at the local escape velocity 
of the disc--halo system, \vesc($r$). 
Therefore, the efficiency of mass ejection depends on mass. 
For halos less massive than $\sim 10^{12}$ \msun, the outflows were found to be able to 
reduce the initial universal baryon fraction by factors $\sim 2$ and more as the mass is 
smaller (c.f. van den Bosch 2002; Benson et al. 2003; DvdB09).
The mass ejected per unit of area and time from disc radius $r$ is given by
\begin{equation}
\dot{\Sigma}_{\rm eject}(r) = \dot{\Sigma}_{\rm SFR}(r)\ \epsilon\ 
\left(\frac{1000\ \kms}{\vesc(r)} \right)^{n},
\label{outflow}
\end{equation}
where $\dot{\Sigma}_{\rm SFR}$ is the SFR surface density, and $\epsilon$
is a free parameter. In the literature, values for $n$ from 1 to 5 were commonly
used in order to reproduce the low--mass luminosity function.
The values $n=2$ and 1 are expected for the 'energy--driven' and 'momentum--driven'
outflows, respectively.  For the former case, assuming that each SN produces an energy 
$E_{\rm SN} = 10^{51}$ erg, and that the the number of SNe per solar mass of stars 
formed is $\eta_{\rm SN} = 0.009 \msun^{-1}$, one obtains namely the 
value of $\sim 1000$ kms$^{-1}$ reported in eq. (\ref{outflow}) and the parameter 
$\epsilon$ can be interpreted as the fraction of SN kinetic energy transferred 
into the outflow. For the latter case ($n=1$), by assuming that each SN produces a 
momentum $p_{\rm SN} = 3\ 10^4\ \msun$kms$^{-1}$ and using the same value of 
$\eta_{\rm SN}$ as above, the normalization velocity in eq. (\ref{outflow}) should
be $\sim 300$ kms$^{-1}$; by keeping 1000 kms$^{-1}$ in eq. (\ref{outflow}), the 
outflow momentum  is already 3.3 times the momentum produced by one SN; 
this should be still multiplied by our $\epsilon$. However, we should say that 
the 'momentum--driven' case is thought for the effect of massive--stars rather 
than for the SN contributions (e.g. Murray, Quataert \& Thompson 2005; 
Oppenheimer \& Dav\'e 2008; Grimes et al. 2009); the total momentum (energy) 
injected by massive stars during their life is much higher than the one injected by SNe.

Because the outflow rate is proportional to the SFR (eq. \ref{outflow}), the outflows
are more efficient in removing gas at earlier epochs, when the SFR is higher. The early 
accreted gas has lower angular momentum than the later accreted gas. Thus, as the
result of disc gas ejection by outflows, the spin parameter $\lambda$ of the 
final disc becomes larger than the spin in case of no outflows, which, by assumption,
is equal to the halo spin parameter (see also DvdB09).

The study of the galaxy--size outflow physics is a complicated hydrodynamic and radiation 
transfer problem, and it requires a very large dynamical range, from parsec to megaparsec 
scales (e.g., Dalla Vecchia \& Schaye 2008; Oppenheimer \& Dav\'e 2008 and more references 
therein). The model described above is certainly an oversimplification, which 
overestimates the feedback efficiency (see for a discussion DvdB09). 
In most of the previous works it was assumed that the ejected gas is lost forever by 
the galaxy--halo system and the proportionality coefficient and exponent in eq. 
(\ref{outflow}) were varied in order to reproduce the low--luminosity side of the 
observed luminosity function as well as galaxy scaling laws like the Tully--Fisher 
relation (c.f. Benson et al. 2003; DvdB09). Feedback--driven outflows seem to be 
also necessary to explain the steep mass--metallicity relation of galaxies at 
redshifts $z\sim 1-2$, and the observed metallicities of the IGM at high redshifts 
(Aguirre et al. 2001; Oppenheimer \& Dav\'e 2006; Finlator \& Dav\'e 2008; DvdB09).

The full modeling of feedback--driven outflows is out of the scope of this paper. 
Instead, we aim to explore in a semi--empirical way the effects of different 
kinds and levels of feedback on the SFR history of disc galaxies as a function of mass. 
The exploration is semi--empirical in the sense that we take care to agree with the 
present--day \mh--\ms\ relation inferred from matching the galaxy stellar
(luminosity) and halo mass cumulative functions. In Fig. \ref{MhMs} we show this 
relation as inferred by different authors (Shankar et al. 2006; Baldry, Glazebrook 
\& Driver 2008; Conroy \& Wechsler 2009; Moster et al. 2009; Rodr\'{\i}guez-Puebla 
et al. 2009, in prep.). Different observed galaxy luminosity functions, different methods 
to pass from luminosity to \ms, and different halo mass functions (analytical fits 
or halos directly from N--body numerical simulations), as well as different 
corrections to account for halo groups and sub-halos, were used by each one of 
these authors. For completeness, we reproduce in Fig. \ref{MhMs} also the more 
direct (but yet very uncertain) inferences based on weak lensing studies for 
late--type galaxies (squares with large vertical error bars; Mandelbaum et al. 2006). 

The \mh--\ms\ relation has been inferred commonly using the whole galaxy population
and all the halos. In the case of Baldry et al. (2008; long dashed blue line), 
the used galaxy sample refers to galaxies in the field, where disc galaxies dominate. 
Since our study refers to disc galaxies, the \mh--\ms\ relation should
be inferred from a stellar mass function for only disc galaxies and from a halo mass function
related to halos that will host disc galaxies. This inference has been performed
in Rodr\'{\i}guez-Puebla et al. (2009), who used the \ms\ function for SDSS late--type (blue) central galaxies
as reported in Yang, Mo \& van den Bosch (2009), and the halo \mh\ function corrected
for halos that did not suffer a major merger (mass ratio larger than 0.2) since $z=0.85$;
Governato et al. (2009) have shown that after a major merger took place before $z\sim 0.8$,
the galaxy can regenerate a significant disc until the present epoch. Rodr\'{\i}guez-Puebla 
et al. checked that the fractions of late--type galaxies and 'quiet' halos with respect
to their corresponding distribution functions were similar, around $55\%$ in both cases.  
The solid red line in Fig. \ref{MhMs}
reproduces the results by Rodr\'{\i}guez-Puebla et al. including the 
1 $\sigma$ uncertainty (dotted red lines which encompass the area of dotted horizontal 
lines); this uncertainty is mainly due to the uncertainties in the population synthesis 
models used to estimate \ms.

We use the \mh--\ms\ relation by Puebla-Rodr\'{\i}guez as indicative, intending 
that the feedback--driven mass ejection shall help reproduce roughly such 
a (semi--empirically derived) relation. It should be stressed that our conclusions below
regarding SSFRs are unchanged if our models are calibrated to reproduce any of the
other \mh--\ms\ relations showed in Fig. \ref{MhMs}.
We focus our study only on galaxies with $\ms \lesssim 10^{10.5} \msun$,
those for which the SSFR-DS phenomenon becomes strong (see \S 2). 

In panel (a) of Fig. \ref{MhMs}, the thick dot--dashed lines show our results for the
cases $n=1$, 2, and 5 in clockwise sense, respectively. In each model the outflow 
efficiency $\epsilon$ is fixed in order to obtain the best agreement with the 
\mh--\ms\ relation at $\ms = 10^{10} \msun$. 
The higher $n$ is, the shallower the \mh--\ms\ relation at low masses;
for example, the slopes in the range $9.5\lesssim$ Log\ms\ $\lesssim 10.5$
are 0.79, 0.62, and 0.36 for $n=1, 2$, and 5, respectively (the corresponding
slope for the Rodr\'{\i}guez-Puebla et al. inference is 0.65). 
 We have experimented also with 
an outflow model with constant mass loading (i.e., $n=0$); as expected, the models follow
a correlation steeper than the case $n=1$. If outflows are not introduced (\fb=0.04=const.
is used), then the correlation is only slightly shallower than the case $\mh\propto \ms$
in Fig. \ref{MhMs} (slope $\approx 0.94)$;
this is because in our models the lower the disc mass is, the less efficient is
the process of gas transformation into stars. 

As seen from Fig. \ref{MhMs}, a reasonable agreement below 
$10^{10.5} \msun$ with the observational inferences is obtained for $n=2$ 
('SN energy--driven' outflows) with a high SN energy transference efficiency, $\epsilon = 0.62$. 
The values of $\epsilon$ for different cases, as well as the obtained values of 
\ms\ at $z=0$ for different halo masses, are presented in Table 1.

\begin{table*}\centering
\begin{tabular}{ccccccccccccccc}
\hline\hline
    &  &  &  &  &  &  \multicolumn{9}{c}{$LogM_h$} \\
$n$ & $\epsilon$ & $Log(m)$ & $\alpha$ & $\mu$ & $\ $ &  {\it 11.2} &  {\it 11.4} &  {\it 11.7} &  {\it 12.1} 
& $\ $ &  {\it 11.2} &  {\it 11.4} &  {\it 11.7} &  {\it 12.1}   \\
\hline
    &  &  &  &  &  &  \multicolumn{4}{c}{$LogM_*$} &  &  \multicolumn{4}{c}{$LogSSFR$} \\
\cline{7-10} \cline{12-15}\\
1 &    1.83     &                          &               &            &    &  9.38 &  9.63 & 10.00 & 10.48 &    & -1.39 & -1.37 & -1.34 & -1.32  \\
2 &    0.62     &                          &               &            &    &  9.22 &  9.54 & 10.00 & 10.60 &    & -1.36 & -1.34 & -1.31 & -1.28  \\
5 &    0.024   &                          &               &            &    &  8.65 &  9.18 & 10.00 & 10.90 &    & -1.30 & -1.26 & -1.19 & -1.20  \\
1 &    7.23     &        10.0           &     1.0     &     0     &    &  9.05 &  9.43 & 10.00 & 10.77 &    & -1.25 & -1.17 & -1.13 & -1.12  \\
1 &    3.18     &        10.5           &     1.0     &     0     &    &  9.20 &  9.51 & 10.00 & 10.70 &    & -1.33 & -1.26 & -1.19 & -1.15  \\
1 &    5.95     &        10.0           &     0.5     &     0     &    &  9.15 &  9.49 & 10.00 & 10.69 &    & -1.28 & -1.22 & -1.18 & -1.17  \\
1 &    9.61     &        10.0           &     1.0     &    -1     &    &  8.73 &  9.15 & 10.00 & 11.08 &    & -1.28 & -1.05 & -0.92 & -1.07  \\
1 &    6.46     &        10.0           &     1.0     &     1     &    &  9.38 &  9.63 & 10.00 & 10.48 &    & -1.44 & -1.37 & -1.34 & -1.35  \\
2 &    2.32     &        10.0           &     1.0     &     1     &    &  9.21 &  9.53 & 10.00 & 10.62 &    & -1.40 & -1.33 & -1.31 & -1.30  \\
\hline\hline
\end{tabular}
\caption{Predicted \ms\ and SSFR at $z=0$ as a function of halo mass \mh\ for different models
with only outflows (lines with columns 3--5 empty) and with both outflows and gas re--accretion. 
All the masses are in units of \msun, and the SSFR is in units of Gyr$^{-1}$.}

\end{table*}

\subsection{Re--accretion of the ejected gas}

The assumption that gas is lost forever is strong and likely unrealistic. 
The outflow propagates along the intra--halo and/or intra--filament media and 
even if gas escapes from the current gravitational potential, it will be slowed 
down and stopped by further interaction with the intergalactic medium, and then 
re--accreted later by the halo--galaxy system
(see e.g., Bertone et al. 2007; Oppenheimer \& Dav\'e 2008). 

 The motion of an ejected disc gas element can be calculated by taking into account the 
halo--galaxy gravitational potential and the viscosity of the surrounding gas. 
Once the ejected gas element stops with respect to the gas around, both the ejected 
gas element and the circungalactic gas will infall together onto the disc galaxy.
This is a rather complex hydrodynamic process, demanding in computing time and, in
any case, uncertain due to the stochastic nature of the problem.
Instead, we adopt here a simple approach that can be easily implemented in our numerical code.
Let us consider that during the time interval elapsed between the ejection of a gas 
element and its re--infall onto the disc, a certain amount \Mso\ of primary cosmological 
gas (see above for a definition) accretes onto the disc. The information related to the 
surrounding gas that breaks the outflow (mostly related to the infalling cosmological gas), 
as well as of the dynamics in the gravitational 
field, are actually summarized in the parameter \Mso\ 
(see for a similar statement Dubois \& Teyssier 2008). In other words, the physics 
concerning to the outflow/re--accretion process can be parametrized through \Mso.  
With this idea in mind, we follow the next simple approach for calculating gas 
re--accretion in our code. For a given disc gas element, as soon as is ejected, 
the acumulated amount of the accreted primary cosmological gas is numerically 
calculated time by time. When such amount reaches the value of \Mso, then the given mass element 
is reintegrated to the disc. This procedure is applied to each ejected gas element
at any time. Adopting such an approach, the physics reduces to fix \Mso\ in a parametric 
fashion. 

Actually we introduce a lognormal distribution around our parameter \Mso\ in 
order to take into account the stochastical nature of the problem.
Such distribution is defined by 
\begin{equation}
dP=\frac{\alpha}{\sqrt{\pi}} \ e^{- \alpha^2 \left( ln\left( \Msop \right) -  
\langle ln \left(\Msop\right) \rangle  \right) ^2}dln\left( \Msop \right), 
\label{lognormal}
\end{equation}
where $\alpha$ regulates the width of the distribution around the logarithmic mean 
value $\langle ln (\Msop) \rangle$; $\alpha$ and \Mso =  $e^{(\langle ln (\Msop) \rangle)}$ 
are introduced as free parameters.

By means of cosmological hydrodynamical simulations, Oppenheimer \& Dav\'e (2008) 
have found that the re--accretion time scales with galaxy mass roughly as 
$M_{\rm gal}^{-1/2}$ and they interpret this to be the case when environmental effects 
are dominating the retardation of outflows. Such a behavior implies in our case that 
\Mso\ should be roughly constant or moderately decreasing with mass\footnote{
The properties of \Mso\ can be estimated also by simple momentum or energy conservation 
arguments. Let us consider a given outflow of mass $M_o$ moving in a given solid angle
at the ejected velocity $V_{o}$ in the halo potential; then, $M_o V_o^{a} = 
(M_o + M_{\rm p})\ V^{a}$, where $M_{\rm p}$ is the mass piled by the outflow, $V$ is 
the shock velocity, and $a=1$ or $2$ for momentum or energy conservation, respectively. 
A rough and general estimate for $V$ 
could be the typical halo circular velocity, $V_c$, which approximately scales with the 
halo mass as $V_c\propto M_h^{1/3}$. Therefore, the mass piled by the outflow is
$M_{\rm p}\approx (M_o V_o)^a/M_h^{a/3}$. Such mass will drag the mass element $M_o$ 
back to the disc. In other words, $M_o$ is reintegrated to the 
disc when, within the given solid angle, the amount of mass $M_{\rm p}$ of primary gas 
is accreted. Thus, \Mso\ is just $M_{\rm p}$ scaled to the entire solid angle $4 \pi$.
From this crude approximation, $\Mso\propto M_h^{-1/3}$ or $M_h^{-2/3}$, dependencies
that are within the range explored below by us.}. This way, for massive galaxies, 
\Mso\ is attained soon and gas re--accretion 
becomes consequently efficient (galactic fountains), while for lower mass galaxies, 
attaining \Mso\ takes longer times; for the smallest galaxies, it may takes eventually 
periods longer than the current Hubble time in which case the gas does not return. Given 
the large uncertainties in the physics of the process, we introduce a parametric 
expression for \Mso:
\begin{equation}
\Mso  = \frac{m}{\msun} \left(\frac{\mh}{2\ 10^{11} \msun} \right)^\mu, 
\label{reinfall}
\end{equation}
where $m$ and $\mu$ are parameters to be probed.

Regarding the angular momentum of the re--accreted gas, the situation is even 
more uncertain. We assume that the specific angular momentum of this gas is a 
fraction $f_j$ of the one corresponding to the currently hierarchically infalling 
cosmic gas.
The results to be presented here depend poorly on the assumed fraction within the 
reasonable range of $f_j\approx 0.3-0.7$ 

In panel (b) of Fig. \ref{MhMs}, the thick black curves show our results for the
outflow model with $n=1$ (see eq. \ref{outflow}) and constant values of the re--accretion
\Mso\ parameter, i.e. with $\mu = 0$ (see eq. \ref{reinfall}): $m = 10^{10}$ (solid line)
and $m = 10^{10.5}$ (dashed line), in both cases using $\alpha = 1$ in the probability
distribution of \Mso\ (eq. \ref{lognormal}); and $m = 10^{10}$ but $\alpha = 0.5$ 
(dot--dashed line). 
The corresponding slopes of the \mh--\ms\ relation in the range $9.5\lesssim$ Log\ms\ $\lesssim 10.5$ are 
0.58, 0.57, and 0.52, close to the slope of the observations.
For outflow models with $n$ larger than 1, after taking into
account re--accretion, the \mh--\ms\ relations become significantly shallower than 
the observational inferences. 
In Table 1 we report our results at $z=0$ (\ms\ and SSFR for different halo masses) 
for the re--accretion parameters mentioned above as well as for some cases with 
$\mu \neq 0$, and for  $n=2$. As expected, by including re--accretion and trying to
reproduce the \mh--\ms\ relation implies very high outflow efficiencies.
We see also that the results depend actually weakly on the dispersion parameter $\alpha$
(e.g., compare rows 4 and 6 from Table 1).

\subsection{Strategy}

A given galaxy model is defined by the halo mass, \mh, its MAH, and its $\lambda_h$. 
The initial baryon fraction is assumed equal to the universal one, \fb=\fbuniv, but
the feedback--driven outflow and gas re--accretion parameters define the actual
value of \fb.   
Since we are interested here in generic evolutionary trends related to the SFR 
and \ms\ histories as a function of mass, we study only the ``central'' models of 
different masses characterized by:
\begin{itemize}
\item  the averaged MAH corresponding to the given halo mass \mh, 
\item a value of $\lambda _h=0.03$, which is somewhat smaller than the mean of 
all relaxed haloes measured in numerical simulations (e.g., Bett et al. 2007).  
\end{itemize}
Note that these physical ingredients have in fact probability density 
distributions (not taken into account here) that, of course,  will produce an  
intrinsic scatter in the values of the galaxy properties studied.

We first study models with galactic outflows only, and then include the
possibility of gas re--accretion. In all the cases the parameters of the 
feedback--driven outflows and gas re--accretion are chosen 
appropriately to reproduce the \mh--\ms\ relation inferred semi--empirically (Fig. 2). 
Our aim is to explore whether the model SSFR histories for different masses 
agree or not in general with the observational trend of SSFR-DS discussed in 
\S 2 (Fig. \ref{obs}).

\section{Results}

The observational data in the SSFR vs. ($1+z$) diagram (Fig. \ref{obs})
refer to curves of constant \ms\ corresponding to galaxy populations at 
different redshifts rather than to curves equivalent to the evolution of 
individual galaxies. However, as discussed in \S 2, at least for $z\lesssim 1$, 
the later curves seem to be only slightly steeper than the former ones in the 
SSFR vs. ($1+z$) diagram. Having in mind this small difference (note that 
at $z\sim 0$, the comparison of models and observations is completely fair), 
we compare below the observational data with results for individual galaxy 
evolution models, which end at $z=0$ with stellar masses similar to those 
corresponding to the data: Log (\ms/\msun) = 9.5, 10.0, and 10.5. The massive
case, Log (\ms/\msun) = 11.0, as has been explained in \S 4, is out of the
scope of the problem studied herein. 
Recall that our models are constrained here to agree with the \mh--\ms\ relation inferred 
semi--empirically (Fig. \ref{MhMs}).

\subsection{Galactic outflows only}

\begin{figure}
\vspace{8.8cm}
\includegraphics{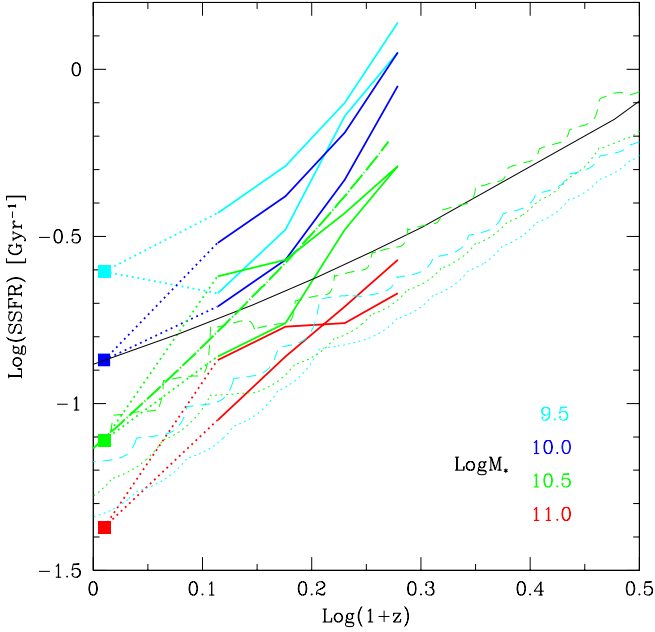} 
\caption{Same as in Fig. \ref{obs} but including evolutionary tracks of individual galaxies.
The dotted curves are for models with outflows and for the favorite case $n=2$; the 
upper green and lower cyan lines are for $\ms = 10^{10.5}$ \msun\ and $\ms = 10^{9.5}$ \msun,
respectively. The dashed curves are for models with outflows ($n = 1$) and gas re--accretion
($\mu = 0, m = 10^{10}, \alpha = 1.0$); the upper green and lower cyan lines are for 
$\ms = 10^{10.5}$ \msun\ and $\ms = 10^{9.5}$ \msun, respectively.
}
\label{evolution}
\end{figure}

We first experiment with the usual ejecting SN feedback model,
where the gas in the outflows (galactic super-winds) is assumed
to be lost forever from the halo (\S\S 4.1). In Fig. \ref{MhMs} we showed
that the semi--empirically derived \mh--\ms\ relation is better
reproduced by the $n=2$ outflow model ('energy--driven' outflows).
The SSFR evolution for this case and for the present--day masses
Log(\ms/\msun) = 9.5 and 10.5 are plotted in Fig. \ref{evolution}
with the dotted cyan (lower curve) and
green (upper curve) lines, respectively. The SSFR tracks corresponding to masses 
in between this mass range lie within these two curves.  

For all the masses, the SSFR increases with $z$ with a trend in rough agreement
with observations, but there is a small but systematical trend of higher 
SSFRs as the mass is larger, {\it opposite to the strong observed SSFR-DS phenomenon}. 
A good fit to the model results from $z=0$ to $z\sim 3$ is: 
\begin{equation}
SSFR = 0.005 \ \ms_{,10}^{0.1} \ (1+z)^{2.2} \ {\rm Gyr}^{-1}.
\label{law}
\end{equation}
Such a behavior is consequence of (i) mainly the CDM halo MAHs which
drive the gas infall process and, consequently, the SF histories: the less
massive the halos, the lower the specific mass aggregation rates
at late epochs (see Fig. \ref{obs}); and (ii), in a less extent, of the fact 
that the disc outflow gas ejection is more efficient for lower mass halos. 
The SSFR histories of low--mass models have significantly over--quenched present--day SFRs, 
i.e. their stellar masses had to be assembled with SFRs in the past higher 
than the current one. In general, all the models since $z\sim 2$ lie below the curve
$1/[t_H$($z$) -- 1 Gyr] (1-R) of approximately constant SFR history 
(thin black solid line in Fig. \ref{obs}; see \S 3).

The most reliable observations are for local galaxy samples. In Fig. \ref{local}
we plot the fit and its 1 $\sigma$ dispersion to the SSFR--\ms\ correlation given 
for local SDSS galaxies by Salim et al. (2007). They report the fit for only
star forming galaxies (solid blue line) and for both this sample and the one of
galaxies with AGN (short--dashed red line). In both cases the galaxy SSFRs 
significantly decrease as \ms\ increases. The three long--dashed magenta curves 
shown in this plot correspond to models with only outflow and
with values of $n = 5$, 2, and 1 from top to bottom, respectively (see also
Table 1). Thick line is used for our preferred $n = 2$ case (see panel (a) of 
Fig. \ref{MhMs}).  The difference between models (weak upsizing) and observations 
(strong downsizing) in the trend of the SSFR--\ms\ relation is significant. 
In particular, for $\ms\lesssim 10^{10.5} \msun$, the lower the mass, the bigger 
the differences between the model SSFRs and those inferred from observations. 
For $\ms> 10^{10.5} \msun$, the differences tend to disappear.

\subsection{Galactic outflows + re--accretion}

In \S\S 4.2 we described our parametric model to account for the re--accretion
of the ejected gas by feedback--driven galactic outflows. The models with
re--accretion agree with the \mh--\ms\ relation only for outflows with $n=1$
(see panel b of Fig. \ref{MhMs}). In Fig. \ref{evolution}, the case with
$n=1$, $\mu = 0$, $m = 10^{10}$ (see eq. \ref{reinfall}) and $\alpha = 1$
(see eq.\ref{lognormal}) is plotted for two models that end 
at $z=0$ with Log(\ms/\msun) = 9.5 (upper dashed cyan line) and 10.5 (lower
dashed green line). We have experimented also with many other re--accretion
parameters (see Table 1).  
As expected, the SSFRs are higher for the models with re--accretion.
This is because the rate of later re--accretion of the ejected gas sums 
up to the hierarchical halo accretion rate, raising the SFR. 
From Fig. \ref{evolution} we see that the SSFR increases with ($1+z$) to a power
not strongly dependent on mass and actually, similar to the case of no re--accretion 
(dotted curves; see eq. [\ref{law}]). 
It is obvious that if re--accretion is included, then the mass loading in the 
outflow model should be increased in order to reproduce the same \mh--\ms\ relation
than without re--accretion.
For example, compare the models given  in rows 1 and 4 in Table 1 ($n=1$ outflow case), 
where $\epsilon$ increased from 1.83 to 7.23; note that the SSFR increased
for all the masses, but more for the more massive models.

The physics of the model is rather complex. For \Mso = const., the larger the mass,  
the shorter the time period of gas re--incorporation into the disc because \Mso\ 
is equaled quickly by the accreted primary cosmological mass 
in more massive galaxies. Furthermore, a given mass element can be ejected and 
re--accreted several times depending on the outflow efficiency.
Given that massive galaxies recover their ejected gas sooner than low mass galaxies, then
the same stellar mass (\mh--\ms\ relation) is reached decreasing the power $n$ of eq. 
(\ref{outflow}), i.e. decreasing the low mass galaxies outflow with respect to massive 
galaxies.  

The behavior of the SSFR with mass is again against the SSFR--DS. 
For the value of \Mso $= 10^{10}$ \msun\ used here, models with 
$\ms\sim  10^{10.5}$ \msun\ raise their SSFR to values slightly lower
on average than observational inferences, while the lowest mass models, only very slightly
raise their SSFR at late epochs, far away from the values required to reproduce
the high SSFRs inferred for low--mass disc galaxies.

Further, we have experimented with cases where $\mu$ in eq. (\ref{reinfall}) is
varied. For $\mu < 0$ and for a fixed value of $m$, the epoch when the SSFR 
becomes higher due to gas re--accretion is delayed with respect to the case 
$\mu = 0$. This increase is early and large for more massive galaxies, but with time 
the SFR--ejection--re--accretion process is stabilized. For small galaxies, 
the influence of re--accretion over the SSFR begins at late epochs. Overall, 
taking as an example the case $\mu = -1$, galaxies that end with 
$\ms\grtsim 10^{10.5}$ \msun\ have
SSFR histories close to those inferred from observations. For smaller galaxies, 
while their SSFRs are now higher than in previous cases, they remain being far from the
observational constraints. Models keep showing a trend in disagreement with the SSFR-DS trend.  
Regarding the cases with $\mu > 0$, the SSFRs of less massive galaxies result even lower
than in the $\mu = 0$ case. 

\begin{figure}
\vspace{8.8cm}
\includegraphics{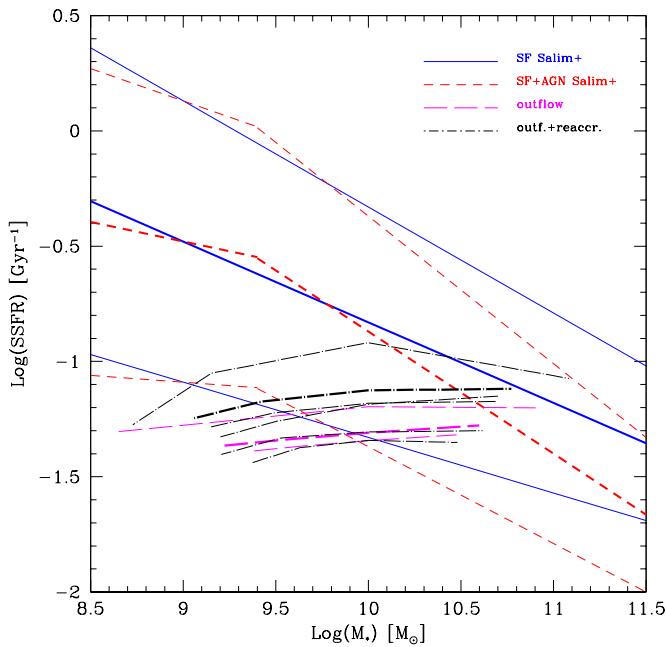} 
\caption{Local ($z\sim 0$) SSFR--\ms\ relation. The fit to the SDDS data  
(average and 1 $\sigma$ scatter) by Salim et al. (2007) is plotted with 
solid blue lines for only star forming galaxies, and with short dashed red lines
for both the star forming and AGN galaxies. The long--dashed magenta curves
are for models with only outflows and $n = 5, 2$ and 1 (respectively from top 
to bottom). Our preferred case, $n = 2$, is highlighted with bold. 
The  dot--long--dashed black curves are for models with outflows and 
gas re--accretion. Our preferred case, $n = 2$, $\mu = 0$, $m = 10^{10}$,
$\alpha = 1.0$ is highlighted with bold. For the other curves, see text and
Table 1. In all the cases, models show a weak 'upsizing' behavior contrary
to the strong 'downsizing' of the observations.
}
\label{local}
\end{figure}

In Fig. \ref{local} the loci of the different re--accretion models in the $z=0$
SSFR--\ms\ diagram are plotted (see also Table 1). The $\mu = 0$, $m = 10^{10}$
fiducial case is shown with the thick dot--dashed black line. The uppermost dot--dashed
curve corresponds to our $\mu = -1$ case. The dot--dashed curve immediately lower 
to the fiducial case is for the same values of this case but for $\alpha = 0.5$.
The next lower curve is as the fiducial case but for $m = 10^{10.5}$. 
Then follow the curves corresponding to our $\mu = 1$ and $n = 2$ cases (see Table 1). 
As seen, for a reasonable range of values for the re--accretion model parameters, 
the SSFR as a function of \ms\ remains in conflict with observations, and the SSFRs of 
low--mass models are too low.

\section{Summary and discussion}

By means of self--consistent models of disc galaxy evolution inside growing \LCDM\ 
halos, the effects of feedback--driven disc ejective outflows (galactic super-winds)
and gas re--accretion on the evolution of the SSFR for low--mass galaxies
($\ms\lesssim 10^{10.5} \msun$ at $z=0$) have been explored. 
We have studied only the 'central' (average) models corresponding to different halo 
masses since our main goal was to probe general trends regarding the evolution of the
SSFR and its dependence on mass.

\subsection{The effect of gas outflows}

The rate of mass ejection was assumed proportional 
to the local SFR and inversely proportional to the local escape velocity \vesc\ to 
a given power $n$ (see eq. \ref{outflow}, for $n=2$ and $n=1$ the outflows are 
expected to be energy-- and momentum--driven, respectively). The parameters of the 
outflow models were explored in the light of the \mh--\ms\ relation at $z=0$ (galaxy 
efficiency) inferred from a semi--empirical approach that makes use of the
observed luminosity (\ms) function (Fig. \ref{MhMs}).

Our best outflow models were chosen to be those that agree with the \mh--\ms\ relation
inferred for disc (blue) galaxies and halos that did not suffer a major merger
since $z = 0.85$. We have found that the slope of the \mh--\ms\ relation at masses
$\ms\lesssim 3\ 10^{10}\ms$ is better attained for the outflow model with $n=2$ 
(energy--driven SN feedback, eq. \ref{outflow}) and for a high SN energy efficiency,
$\epsilon =0.62$ (Fig. \ref{MhMs}, panel a).

For the $n=2$ outflow case, the SSFR of all masses raises with 
redshift proportional to ($1+z$)$^{2.2}$ up to $z\sim 3$ with a little 
dependence on (present--day) mass, as $\ms^{0.1}$ (eq. \ref{law}).
The SSFRs at low redshifts for models with $\ms\lesssim 10^{10.5} \msun$  
are well below the constant SFR curve $1/[t_H$($z$) -- 1 Gyr](1-R) (see \S 3), 
which means quiescent late SF activity, and miserably fail in reproducing 
the observational inferences (Fig. \ref{evolution}). In particular, the models 
show a trend contrary to the observed SSFR-DS phenomenon. Such a conflict is
clearly seen in the SSFR vs \ms\ diagram for local galaxies ($z\sim 0$), where
the observational data are more reliable: while the SSFR of observed galaxies 
significantly decreases with \ms, the models show a slight increase with \ms.
The situation is similar for model outflows with $n=1$ and $n=5$ (Fig. \ref{local}). 

We stress that the conflict is not related to the observed low SSFRs of massive disc 
galaxies ($\grtsim 10^{10.5} \msun$) with respect to models, but instead is related 
to {\it (i) the observed trend of SSFR increasing as the mass decreases (SSFR--DS), 
and (ii) to the too high values of the SSFRs of galaxies with masses 
$\ms < 10^{10} \msun$, both items applying since $z\sim 1$.} 

In the models, on one hand, the SFR history is largely driven by the hierarchical 
halo mass aggregation rate\footnote{In our models, the SFR history depends also 
on the local disc gas surface density and on some disc dynamical and hydrodynamical 
properties. Nevertheless, the dominant factor in the SFR histories of our modeled disc 
galaxies is the cosmological gas infall rate.}, which on average increases with 
redshift at nearly the same rate for all masses, but at a given $z$, massive halos have 
rates slightly higher than those of less massive halos (see Fig. \ref{obs} and eq. \ref{SSFRdm}). 
On the other hand, the ejective feedback scheme used here produces proportionally more 
gas loss (hence, less later SF) for the lower mass discs than for the massive ones.   
We conclude that the low SSFRs of low--mass disc galaxy models with outflows, 
and the (slight) increase of such SSFRs with \ms, opposed to the observed 
strong SSFR--DS phenomenon --the lower the mass, the higher the SSFR, 
{\it is a natural consequence of the \LCDM\ halo assembling and, in a minor level, of the feedback--driven outflow schemes commonly used to reproduce the local \mh--\ms\ relation.}

\subsection{The contribution of gas re--accretion}

We further explored the possibility that the gas ejected from the disc--halo system 
can be re--accreted later onto the disc. Our scheme for such a process is very 
general and encompasses a large range of possibilities for the unknown and complex 
outflow hydrodynamic and radiation transfer processes. Instead of introducing as a 
key parameter the time elapsed between when a given mass element is ejected
and it is further re--accreted,
we use as a parameter the primary cosmological gas mass that will be accreted 
during this period, \Mso. This way, we take into account in some way the amount of 
circumgalactic mass that interacts with the outflow, amount of mass that in our 
models is related to the MAH of the given halo. We use a parametrization for 
\Mso\ given by eq. (\ref{reinfall}), and a lognormal distribution (eq. \ref{lognormal})
for such a parameter is considered in order to introduce a (natural) dispersion 
around the re--accretion times of the ejected mass shell.

Our results show that the gas re--accretion may raise the SSFR of disc galaxies 
but it does it in the incorrect direction regarding mass: while for low mass galaxies 
the increase in SSFR is small, for the massive galaxies the increase is significant at
all epochs (Fig. \ref{evolution}). 
We have experimented with several cases: \Mso\ constant for all masses, 
and \Mso\ decreasing or increasing with mass. The decrease of SSFR as \ms\ decreases
(upsizing) is seen in all the cases (for $z\sim 0$, see Fig. \ref{local}).

 We conclude that for models of disc galaxy formation and evolution in the context
 of the \LCDM\ cosmogony, the problem of the SSFR--DS is not solved by an 
interplay of outflows and re--accretion of gas.

\subsection{Is the SSFR--DS phenomenon well established?}   
   
We have shown in a clear and transparent way the difficulty galaxy evolution models 
in the context of the \LCDM\ cosmology have for explaining the SSFR--DS phenomenon, 
related mainly to sub--$L_*$, star--forming (late--type) galaxies. 
Before discussing some caveats of the models and 
possible cures to the problem, it should be stressed that {\it at the level of
observational inferences there are still large uncertainties.}

The use of SPS models to fit observational data and infer \ms\ and SFR should be
taken still with caution due mainly to the uncertainties in 
stellar evolution (for example in the thermally--pulsating asymptotic giant branch, 
TP-AGB, and horizontal branch phases) and to our poor knowledge of the initial mass 
function, IMF, as well as due to 
degeneracies like the one between age and metallicity (see for recent extensive discussions 
Maraston et al. 2006; Bruzual 2007; Tonini et al. 2009; Conroy, Gunn \& White 2009a; 
Conroy, White \& Gunn 2009b). For example, Conroy et al. (2009a) estimate that stellar
evolution uncertainties can introduce up to $\sim 0.3$ dex statistical error in \ms\ at 
$z\sim 0$. Regarding systematical errors, they are more difficult to evaluate. Maraston
et al. (2006) claim that the stellar masses of galaxies with dominating stellar populations of 
$\sim 1$ Gyr age could be on average $60\%$ lower if their assumptions for 
convective overshooting during the TP--AGB phases are used. It is not easy to estimate
the direction in which these statistical and systematical uncertainties could influence 
the SSFR--\ms\ dependencies at different redshifts; at least, it is not obvious that
the SSFR--DS phenomenon could be eliminated. 

It is healthy to stress that the approach we follow for comparing models and observations
is the recommended one in the sense of minimizing uncertainties. Conroy et al. (2009b) 
show that it is better to first infer the physical properties of observed galaxies (e.g., 
stellar mass and SFR) by using the SPS technique, and then compare them to galaxy evolution 
models, rather than applying the SPS technique to models in order to compare 
their predictions with the direct observables.
 
Other significant sources of uncertainty in the inferred SSFR--\ms\ relations at different 
redshifts are the selection effects due to the incompleteness of the sample, dust absorption, 
environmental effects, and/or limit detections of the tracers of SF due to flux--limits or 
low emission--line signal--to--noise ratios; in addition, obscured AGN emission could be 
contaminating the infrared flux and some of the optical lines used to estimate SFR 
(e.g., Daddi et al. 2007; Chen et al. 2009).

The main concern regarding the SSFR--DS phenomenon among these issues is that 
{\it selection and environmental effects} could bias the observed trend that the SSFR increases
as \ms\ is smaller. For example, 
the high SSFRs of low--mass galaxies could be due to transient star bursts if 
the SF regime of these galaxies is dominated by episodic
processes (external like mergers or internal like statistical fluctuations in massive 
star formation); if among the low--mass galaxies, those with low--SFRs are missed due 
to detection limits, then the SSFRs of low--mass galaxies will be biased on average toward
higher values of SSFR, a bias that increases for samples at higher redshifts.
Nevertheless, in most of the observational studies reporting the SSFR--DS
phenomenon the authors discuss that, while this is possible at some level, hardly 
would it be the dominant factor in the SSFR--DS phenomenon observed
since $z\sim 1$ (see for example Noeske et al. 2007a).

On the other hand, the analysis of very local surveys certainly helps to disentangle 
whether episodic star bursting events dominate or not the SF history of low--mass disc 
galaxies.  From a study of the SF activity of galaxies within the 11 Mpc Local Volume,
Lee et al. (2007; see also Bothwell, Kennicutt \& Lee 2009) have found that 
intermediate--luminosity disc galaxies 
($-19 \lesssim M_B \lesssim -15$ or  $50\lesssim \vm/\kms \lesssim 120$) 
show relatively low scatter in their SF activity, implying factors 2--3
fluctuations in their SFRs; above $\vm\approx 120$ km/s the sequence turns off toward 
lower levels of SSFRs and larger bulge--to--disc ratios. These results are for nearby 
galaxies, where selection effects are minimal, and imply that the SSFRs of disc
galaxies with $\ms\grtsim 5\ 10^{8}$ \msun\ follow a relatively tight sequence, without
strong fluctuations.  
For galaxies smaller than $\vm\sim 50$ km/s (dwarfs) the situation seems different.
The results by Lee et al. (2007) show that a significant fraction of such galaxies are 
undergoing strong episodic SF fluctuations due to the large scatter in their SSFRs. 
Another observational study of nearby galaxies by James et al. (2008), concluded also that 
there is little evidence in their sample of predominantly isolated field galaxies of significant
SF through brief but intense star-burst phases. Therefore, it seems that the tight sequence 
found for normal star--forming (disc) galaxies in the SSFR--\ms\ plane in large surveys as SDSS 
(Brinchman et al. 2004; Salim et al. 2007; Schiminovich et al. 2007) is intrinsic and
due to a high degree of temporal self--regulated SF within individual galaxies. This 
sequence (called the 'main sequence' in Noeske et al. 2007a) seems to persist back to 
redshifts $z\sim 1$ as discussed above.

\subsection{Outlook}

If the SSFR--DS phenomenon is definitively confirmed by observations, then, as we have argued, 
it poses a serious difficulty for current {\it disc} galaxy evolution models in the context of 
the hierarchical \LCDM\ scenario. This difficulty, at one or another level, is present also 
in other galaxy 
evolution approaches, e.g., the SAMs (Somerville et al. 2008; Fontanot et al. 2009; Lo Faro et al.
2009). These models show, more from a statistical point of view than
at the galaxy individual evolution level, that the population of relatively small galaxies 
($\ms\approx 10^9-10^{10.5}$ \msun) is mostly assembled at higher redshifts becoming older, redder,
and with (much) lower SSFRs at later epochs than the observed galaxies in the same mass range.

Our models clearly show that less massive galaxies assemble their stars early because
their dark halos assemble early (earlier on average than more massive ones), and that 
SN--driven outflows, which are more efficient as less massive is the galaxy, contribute to quench
later SF. Gas re--accretion helps to increase moderately the SSFR, but not enough as to
agree with observations for low--mass galaxies. 
The SF--feedback efficiencies used here to produce galactic outflows able to recover the 
required \mh--\ms\ relation are too large according to hydrodynamical simulations 
(c.f. Oppenheimer \& Dav\'e 2008; Dubois \& Teyssier 2008). 
Besides, it should be taken into account that a significant fraction of the SF--feedback 
energy is actually dissipated into the disc ISM turbulence. Concerning the SF physics in our
models, it should be said that the SF is assumed stationary (self--regulated) and related
only to disc internal processes (isolated galaxy).  Most of the model predictions 
(dynamics, structure, gas fractions, etc.) describe well present--day {\it normal disk} galaxies, 
but we stress that we are not modeling, for example, interaction--induced SF and/or SF in a 
bursting (non--stationary) regime.

 The delay of SF activity apparent from the SSFR--DS trends, could be produced by external 
effects; for example large--scale 
gas pre--heating or by the introduction of new physical ingredients in the intra--halo 
medium hydrodynamics and gas cooling process.  
Finally, it could be that the SSFR--DS problem, more than a physical phenomenon related 
to the cosmic gas accretion, is a manifestation of some local process regarding
star formation in environments that change with the galaxy mass and cosmic time and/or 
of a varying stellar initial mass function.

\section*{Acknowledgments}

We are grateful to the referee, Dr. B. Oppenheimer, for his thoughtful comments and suggestions on our
paper, which largely improved its presentation.  
V.A-R. thanks PAPIIT-UNAM grant IN114509 and CONACyT grant 60354 for partial funding. 
A.R-P. acknowledges a graduate student fellowship provided by CONACyT.

\end{document}